\documentclass[10pt,conference]{IEEEtran}
\IEEEoverridecommandlockouts
\usepackage{amsmath,amssymb,amsfonts}
\usepackage{algorithmic}
\usepackage{graphicx}
\usepackage{textcomp}
\usepackage{xcolor}
\usepackage{multirow}
\usepackage{caption}
\usepackage{subcaption}
\usepackage{hyperref}
\usepackage{url}
\usepackage{dblfloatfix}

\usepackage[
backend=biber,
style=numeric,
sorting=none
]{biblatex}
\def\BibTeX{{\rm B\kern-.05em{\sc i\kern-.025em b}\kern-.08em
    T\kern-.1667em\lower.7ex\hbox{E}\kern-.125emX}}

\begin{document}
\sloppy

\title{Near Lossless Time Series Data Compression Methods using Statistics and Deviation\\
}

\author{{Vidhi Agrawal$^{1}$, Gajraj Kuldeep$^{2}$, Dhananjoy Dey$^{1}$}\\
\IEEEauthorblockA{\textit{$^{1}$Indian Institute of Information Technology} 
\hspace{1cm} \textit{$^{2}$Aarhus University}\\
Lucknow, India \hspace{5cm} Aarhus, Denmark\\
\{vidhi.0206, gajrajkuldeep, ddey06\}@gmail.com}  

}

\maketitle

\sloppy

\begin{abstract}
The last two decades have seen tremendous growth in data collections because of the realization of recent technologies, including the internet of things (IoT), E-Health, industrial IoT 4.0, autonomous vehicles, etc. The challenge of data transmission and storage can be handled by utilizing state-of-the-art data compression methods. Recent data compression methods are proposed using deep learning methods, which perform better than conventional methods. However, these methods require a lot of data and resources for training. Furthermore,  it is difficult to materialize these deep learning-based solutions on IoT devices due to the resource-constrained nature of IoT devices. In this paper, we propose lightweight data compression methods based on data statistics and deviation. The proposed method performs better than the deep learning method in terms of compression ratio (CR). We simulate and compare the proposed data compression methods for various time series signals, e.g., accelerometer, gas sensor, gyroscope, electrical power consumption, etc. In particular, it is observed that the proposed method achieves 250.8\%, 94.3\%, and 205\% higher CR than the deep learning method for the GYS, Gactive, and ACM datasets, respectively. The code and data are available at \url{ https://github.com/vidhi0206/data-compression}. 

\end{abstract}

\section{Introduction}
% In recent years, IoT has become an integral part of modern digital ecosystem. The ever-increasing scale of the network leads to the inclusion of a variety of sensor and IoT device nodes. The addition of these nodes increases the already vast amount of data captured by the devices in the IoT network. Storage, transmission, and processing of such high volumes of data pose many potential problems for the environment, users, and developers using resource-constrained devices. A few of the problems are the reduction of the sensor's battery life, the requirement for high processing power and data transmission bandwidth. Radio transmission is power-hungry by its very nature. The power consumption of modern CPUs and micro-controllers, while getting lower with every generation, is still too high. While data processing consumes significantly less energy, data transmission in IoT sensor nodes is very expensive and uses a significant amount of energy. Approximately 80\% \cite{9118600} of the sensor's battery is used up in data transmission.

In recent years, the IoT has become an integral part of the modern digital ecosystem. The ever-increasing scale of the network leads to the inclusion of a variety of sensor and IoT devices. This requires a vast amount of data sensing and transmission by IoT devices. The most common approach to dealing with the data explosion problem in IoT infrastructure is to find the best way to represent, store, and organise information. In order to store and transmit data, different data compression methods have been explored in the literature. Compression methods provide an efficient representation of data, thus decreasing the cost of transmission while maintaining consistency and correctness in the data. This work aims to provide a unique combination of statistics and deviation transformations, and entropy compression techniques that have a high compression ratio.

Some data compression techniques are dedicated to IoT networks, while others are meant for offline compression. Several works on lossless and lossy compression of floating-point time series data have been published. On multidimensional datasets, specialised compressors such as FPZIP \cite{4015488} outperform general-purpose compressors for lossless compression of floating point data. Several lossy compressors, which allow significantly higher compression ratios at the cost of an acceptable level of distortion being present in the data, have also been proposed. Recently, some lossy compressors have been proposed and used. These allow for a better compression ratio with a level of signal distortion that is acceptable. Some of these compressors are LFZip \cite{9105816}, critical aperture (CA) \cite{7076402}, SZ \cite{7516069}, ISABELA \cite{ISABELLA}, NUMARCK \cite{7013047}, etc. Next, we discuss data compression methods, which are directly related to our work.

LFZip \cite{9105816} is a lossy prediction-based compressor for time series data. It uses a prediction-quantization-entropy coder framework with a normalized least mean square (NLMS) or neural network based predictor. It works under a maximum error distortion metric, where the maximum allowable error is user-specified. CA \cite{7076402} retains only a subset of points in time series based on whether the point falls outside the maximum error constraint. It uses linear interpolation for decompression. SZ \cite{7516069} uses curve-fitting models to predict the next point in time series and then quantize the prediction error. Proposed methods in \cite{9105816} have high GPU requirements for training their models. Additionally, trained models may not be implementable on resource-constrained IoT devices. Moreover, this tends to increase computation cost and decrease battery life. The general approach of prediction-quantization-entropy encoding has been extensively used for lossy compression in domain of speech \cite{1890}, images \cite{webp} and videos \cite{6031852}. For time-series data compression, LFZip, CA, and SZ  outperform the other data compression methods. Therefore, in  this paper, we compare simulation results with these methods.

\subsection{Our Contribution}
Our contributions can be summarised as follows.
\begin{itemize}
    \item We propose a two-level compressor for numerical time series data based on statistics and deviation transformations and an entropy encoding framework.
    \item The proposed method archives higher compression ratio as compared to LFZip, CA, and SZ at low computation. 
\end{itemize}

% \subsection{Novelty}
% The proposed algorithm is based on statistical measurements(mean, mode,median and variance) and is therefore computationally lightweight. It has no requirements for excess computational power (GPU) at any point of time, thereby more suitable for resource-constrained devices. The proposed methodology has two level compression that can be performed on same device as well as on different devices suiting to the users need. 

\subsection{Paper Organization}
The paper is organized as follows: Section \ref{Method} describes the proposed compression methods. Section \ref{Results} contains the simulation results. Finally, section \ref{conclusion} concludes the paper.

\section{Proposed Methods}
\label{Method}
In this section, we first describe each step used in the proposed compression methods. Then, steps in decompression are described for the proposed methods.
\subsection{Compression}
\begin{figure}[t!]
\centering
  \includegraphics[width=.5\textwidth]{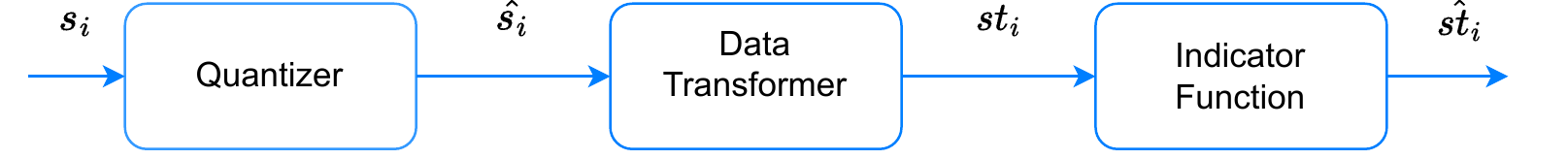}
  \caption{Data transformation framework}
  \label{transformation-flowchart}
\end{figure}

The proposed compression methods use two stages: the first is data transformation, which is shown in Fig. \ref{transformation-flowchart}, and the second is entropy coding.

The data transformation framework contains first step as quantization. Then, statistics of the quantized data is performed using \emph{mode}, i.e., the sample value that appears most often in a block of data. If the frequency of mode value is higher than the certain value $\tau$, mode is subtracted from all the values in the block. Otherwise, difference coding, i.e., the difference between the current sample and the previous sample \cite{xia2014ddelta}, is applied to the block. This step gives deviated values either from the mode or the previous sample in a block. Finally, the indicator function is used to collect zero deviation from this transformed data block. Next, steps in the data transformation framework are explained in detail for time-series data.

% This method is designed for time-series numerical signals. The data compression scheme is a combination of difference coding \cite{xia2014ddelta} and statistical measurements. The method is designed, taking \emph{mode} as the statistical measurement of the data block and the difference between the current and previous/mode reading. This method can be used for both lossy as well as lossless compression by choosing to apply quantisation or not. 

%\subsection{Data transformation Algorithm}

% The proposed statistical algorithm is based on statistical measurements i.e. mean, mode, median and variance. It calculates the mode of the block of data and if the frequency of mode is sufficiently high, mode is subtracted from every value in the block. For low frequency, depending on the value which will provide least variance mean or median is calculated.  The calculated value is then subtracted from the values in the data block.The subtracted value is then added in the beginning of the data block. Post this indicator function is applied as shown in section \ref{IFZ}. This completes the data transformation step. Further enhancements taking this statistical transformation algorithm as the base version are described in the following paragraphs.

The device will collect the time-series signal up to a length \emph{L}. This \emph{L} is to be selected by the user, is a power of 2 greater than 8. This block is then processed by the encoder. The block size can be fixed or selected based on the bit-width of the samples. It is to be tailored to the availability of data, storage, and computational ability of the resource-constrained device. The signal block at $i^{th}$ time will be denoted by $s_i$ and individual values in signal by $x_1, \ldots, x_L$, in the following manner.

%\begin{flalign*}
   $$ s_1 = \;x_1,\; x_2,\; \ldots ,\; x_L$$
%\end{flalign*}

The data transformer processes one data block at a time and consists of 3 stages: quantization, transformation, and indicator function, as shown in Fig. \ref{transformation-flowchart}. We propose two compression methods, namely statistical method \emph{version 1} and statistical method \emph{version 2} based on the data transformation framework. For the proposed methods, the quantization step is the same, but the transformation and indicator function steps differ.

\subsubsection{\textbf{Quantization}}
\label{quantization}
This is a voluntary step for the user and can be chosen to skip in the case of lossless compression.  The floating-point data readings are quantized in this step so as to satisfy the maximum absolute error ($\epsilon$) constraint. This stage eliminates any unnecessary floating point digits from $s_i$. We are using rounding as the quantization approach, in which the fractional portion of the reading is rounded off to the user's desired number of digits. The output signal of the step is represented as $\hat s_i$. Let $\Delta_i= x_i-\hat x_i$ represent the difference between initial value and quantized output. Uniform quantization guarantees that $|\Delta| \leq \epsilon$, which implies that $|x_i-\hat x_i| \leq \epsilon$. For example, $x_1 = 124.3472$ and
user wants current reading to be rounded-off to nearest two decimal places then,
$\hat x_1 =124.35$.

\medskip

\subsubsection{\textbf{Transformation and Indicator Function}}
\label{transformation}
Transformation uses mode or difference to reduce the average number of digits present in the value that occurs in a data block. The indicator function is the post-transformation step; it finds the redundant values, i.e., zeros, found in the block. This step removes the zeros from the $st_i$ by using an indicator function (IFZ). This allows us to remove the redundant zero values from the block, thereby decreasing its size while only adding a single value, i.e., the IFZ value. The IFZ value represents the zero and non-zero values present in the block using the bitwise representation. The output of this step is represented as $\hat {st_i}$ for input $st_i$.

The transformation and indicator function steps for version 1 and version 2 are described below: 

% Mode($mod_i$) and frequency of $mod_i$ for the block $\hat{s_i}$ is found. If the frequency of mode $< \tau$ in this, then difference coding is selected otherwise statistical encoding.

\medskip
\noindent
\textbf{Version 1: }
$mod_i$ is the most often repeated value in the $i^{th}$ data block. If the frequency of mode $< \tau$ in this, then difference coding is selected otherwise statistical encoding.
\begin{itemize}
    \item \textbf{Frequency of mode $> \tau$:}  In this case, 1 and $mod_i$ are prepended to the beginning of the signal data and then $mod_i$ is subtracted from every reading in the signal. The transformation for the first small block, $s_i$, is given as,
    \[ st_1 = \;1,\; mod_1,\; x_1 - mod_1,\; x_2 - mod_1,\; \ldots,\; x_L - mod_1, \] and the output of the indicator function is given as:
     \[\hat {st_i} =\;1,\; mod_1,\; IFZ,\; nonzero \;values.\]
    \item  \textbf{Frequency of mode $< \tau$:} In this case, 0 is prepended to the beginning of the signal data and previous reading is subtracted from the current reading. The transformation of first block, $s_1$, is given as,
    \[ st_1 =\; 0,\; x_1,\; x_2 - x_1,\; x_2 - x_3,\; \ldots,\; x_L - x_{L-1}, \] and the output of the indicator function is given as:
    \[\hat {st_i} = st_i.\]
\end{itemize}

\noindent
\textbf{Version 2}
\begin{itemize}
    \item \textbf{Frequency of mode $> \tau$:}  In this case, $mod_i$ is prepended to the beginning of the signal data and then $mod_i$ is subtracted from every reading in the signal. The transformation for the first small block, $s_i$, is given as,
    \[ st_1 = \; mod_1,\; x_1 - mod_1,\; x_2 - mod_1,\; \ldots,\; x_L - mod_1, \] and the output of the indicator function is given as:
    \[\hat {st_i} =\; mod_1,\; IFZ,\; nonzero\;values.\]
    \item  \textbf{Frequency of mode $< \tau$:} In this case, $x_1$ is prepended to the beginning of the signal data and previous reading is subtracted from the current reading. The transformation of first block, $s_1$, is given as,
    \[ st_1 =\; x_1,\; x_1,\; x_2 - x_1,\; x_2 - x_3,\; \ldots,\; x_L - x_{L-1}, \] and the output of the indicator function is given as:
    \[\hat {st_i} = \;x_1,\;IFZ, \;nonzero\;values. \]
\end{itemize}
Using mode as the statistical measurement allows us to reduce the redundant values with maximum frequency to zero while only adding a number to the block.
\medskip

% \subsubsection{\textbf{Indicator Function}}
% \label{IFZ}
% Post transformation step many redundant values(zeros) can be found in the block. This step removes the zeros from the $st_i$ by using an indicator function (IFZ). This allows us to remove the redundant zero values from the block, thereby decreasing its size while only adding a single value i.e. IFZ value. IFZ value represents the zero and non-zero value present in the block, using the bitwise representation. The output of this step is represented as $\hat {st_i}$ for input $st_i$. 

% \medskip
% \textbf{Version 1}
% \begin{itemize}
%     \item \textbf{Frequency of mode} $> \tau$ : 
    
%     \[\hat {st_i} =\;1,\; mod_1,\; IFZ,\; nonzerosvalues.\]
%     \item \textbf{Frequency of mode} $< \tau$ : 
%     \[\hat {st_i} = st_i\]
% \end{itemize}

% \medskip
% \textbf{Version 2}
% \begin{itemize}
%     \item \textbf{Frequency of mode} $> \tau$ : 
    
%     \[\hat {st_i} =\; mod_1,\; IFZ,\; nonzerosvalues.\]
%     \item \textbf{Frequency of mode} $< \tau$ : 
%     \[\hat {st_i} = \;x_1,\;IFZ, \;nonzerosvalues \]
% \end{itemize}

\begin{flushleft}
For example:\[ st_1 = 1,\; 1290,\; 10,\; -2,\; 0,\; 0,\; 0,\; -1,\; 2,\; 3,\; 0,\; 1,\; 0,\; 0,\; 3,\] 
\hspace{1.1cm}$4,\; 0,\; 1$
\end{flushleft}

The indicator function is applied for last L entries. In this particular case, IFZ is 1, 1, 0, 0, 0, 1, 1, 1, 0, 1, 0, 0, 1, 1, 0, 1, zeros are indicated with 0, non-zero values are indicated by 1. IFZ has 1100011101001101 as binary value and when converted to integer has value of 51021.
\[\hat {st_1} =\; 1,\; 12908,\; 51021,\; 10,\; -2,\; -1,\; 2,\; 3,\; 1,\; 3,\; 4,\; 1 \]

\subsubsection{Entropy Encoding}

The final stage of compression involves applying an universal lossless entropy encoder to the transformed time series data $\hat{st_i}$ to obtain a variable length bit stream. This algorithm uses adaptive arithmetic \cite{adap_arithmatic} or Huffman encoding schemes (static and adaptive) \cite{4051119} \cite{faller1973adaptive} \cite{1055959} \cite{KNUTH1985163}. The encoding takes place at a gateway to further compress the data stream. Here, two or more similar data streams from the same or different device are combined and compressed using an entropy encoder to provide better compression ratio. The implementation of entropy encoding algorithms are taken from the following Github repositories \cite{static_huffman} \cite{dynamic_huffman} and \cite{ada-arithmetic}. Note that, the entropy encoding stage is the same for both version 1 and 2, i.e., we simulate results using both Huffman and arithmetic, but we chose adaptive arithmetic coding for comparison, because it performs better than the other entropy coding in our simulations.

\subsection{Decompression}
The decompression method is similar but occurs in the reverse order of the whole compression scheme. First, entropy encoded bit stream is decoded into collection of data blocks. Various data blocks from the data stream are then separated to obtain individual transformed data blocks $\hat{st_i}$.The reconstruction of transformed data $\hat{st_i}$ then takes place one-step at a time. 

\medskip
\noindent
\textbf{Version 1}
\\
The transformed data has the following structure.
\[\hat {st_i} =\;1,\; mod_1,\; IFZ,\; nonzero\;values.\]
\begin{center}
    or
\end{center}
\[ \hat {st_i}=\; 0,\; x_1,\; x_2 - x_1,\; x_2 - x_3,\; \ldots,\; x_L - x_{L-1} \]
From the transformed data $\hat{st_i}$, a flag integer is input first which tells us whether statistical decoding or difference decoding should take place.
\begin{itemize}
    \item \textbf{flag = 0:} next $L$ numbers are taken and decoded as per procedure of difference coding. The previous decoded value is added to the current encoded one to get the current decoded value.
    \item \textbf{flag = 1:} next 2 numbers are taken. First number is the $mod_i$, second is IFZ. While bitwise iterating the bits of IFZ, whenever we encounter 1 we input a number otherwise we add 0 to the output sequence $\hat st_i$. Now, we have a sequence that is equal to $st_i$. Post this, we will add $mod_i$ to last $L$ numbers in the sequence to recover the original data block $\hat{s_i}$ that we encoded.
\end{itemize}

\medskip
\noindent
\textbf{Version 2}
\\
The transformed data has the following structure.
\[\hat {st_i} =\; mod_1,\; IFZ,\; nonzero\;values.\]
\begin{center}
    or
\end{center}
\[\hat {st_i} =\; x_1,\; IFZ,\; nonzero\;values.\]
From the transformed data $\hat{st_i}$, first three integers($\hat{x_1},\hat{x_2},\hat{x_3}$) are taken as input. If first and 3rd integers are equal, then difference decoding should take place else statistical decoding should take place.
\begin{itemize}
    \item  $\hat{x_1} = \hat{x_3}$:   $\hat{x_2}$ is IFZ number. While bitwise iterating the bits of IFZ, whenever we encounter 1 we input a number otherwise we add 0 to the output sequence $\hat st_i$. Now, we have a sequence that is equal to $st_i$. Post this, we will add $\hat{x_1}$ to last $L$ numbers in the sequence to recover the original data block $\hat{s_i}$ that we encoded.
    \item  $\hat{x_1} \neq \hat{x_3}$:  $\hat{x_2}$ is IFZ number. While bitwise iterating the bits of IFZ, whenever we encounter 1 we input a number otherwise we add 0 to the output sequence $\hat st_i$. Now, we have a sequence that is equal to $st_i$. Post this, we will add $mod_i$ to last $L$ numbers in the sequence to recover the original data block $\hat{s_i}$ that we encoded.
\end{itemize}

\begin{table}[t!]
\centering

\medskip
\caption{Datasets used for evaluation.}
\label{table1}
	\resizebox{.7\columnwidth}{!}{
\begin{tabular}{ l  l  l } 
 \hline
Name & Description & Length\\
\hline
BVP &  Blood volume pulse\cite{wesad} & 3.08 MB\\
EDA &  Electrodermal Activity Sensor \cite{wesad} & 0.27 MB\\
ACM &   Smartwatch accelerometer\cite{10.1145/2809695.2809718} & 16.93 MB\\
GYS &  Smartwatch gyroscope\cite{10.1145/2809695.2809718}  & 17.03 MB\\
GAS & Home activity monitoring  \cite{HUERTA2016169} & 8.26 MB\\
Gactive & Power consumption - active power\cite{power} & 13.89 MB\\
\hline
\end{tabular}}

\end{table}

\begin{table}[htbp!]

\medskip
\caption{\small Comparison of CR for various entropy coding with maximum error of $10^{-3}$.}

\label{table2}

\begin{subtable}[h]{0.50\textwidth}
\centering
\resizebox{.7\columnwidth}{!}{
    \begin{tabular}{| p{1cm} | p{1.55cm} | p{1.55cm} | p{1.6cm} |} 
    \hline
    \multirow{2}{*}{Datasets} & \multicolumn{3}{|c|}{entropy encoding}\\ 
    \cline{2-4} & Static Huffman & Adaptive Huffman & Adaptive arithmetic\\
    \hline
    BVP &  2.78 & 2.75 & \textbf{2.80}\\
    \hline
    EDA &  \textbf{12.65} & 12.62 & 12.59\\
    \hline
    ACM &  5.38 & 5.35 & \textbf{5.44}\\
    \hline
    GYS &  7.68  & 7.66 & \textbf{7.82}\\
    \hline
    GAS & 13.67 & 13.61 & \textbf{13.74}\\
    \hline
    Gactive & 4.14 & 4.12 & \textbf{4.17}\\
    \hline
\end{tabular}}
\caption{Version 1}
\label{table2:v1}
\end{subtable}

\begin{subtable}[h]{0.50\textwidth}
    \centering
    \resizebox{.7\columnwidth}{!}{
    \begin{tabular}{| p{1cm} | p{1.55cm} | p{1.55cm} | p{1.6cm} |} 
     \hline
    \multirow{2}{*}{Datasets} & \multicolumn{3}{|c|}{entropy encoding}\\ \cline{2-4} & Static Huffman & Adaptive Huffman & Adaptive arithmetic\\
    \hline
BVP &  2.42 & 2.41 & \textbf{2.44}\\
\hline
EDA &  \textbf{12.78} & 12.75 & 12.75\\
\hline
ACM &  4.60 & 4.58 & \textbf{4.64}\\
\hline
GYS &  7.03  & 7.01 & \textbf{7.13}\\
\hline
GAS & 15.02 & 14.94 & \textbf{15.10}\\
\hline
Gactive & 4.16 & 4.14 & \textbf{4.18}\\
\hline
\end{tabular}}
\caption{Version 2}
\label{table2:v2}
\end{subtable}
\end{table}

\section{Simulation Results}
\label{Results}
In this section, first we describe the datasets, computational requirements and performance metrics. Then, we evaluate the performance of the proposed compression methods. Finally, we compare the proposed methods with the state-of-the-art methods.
\subsection{Experimental setup}

We simulate the proposed data compression methods, i.e., version 1 and version 2, on several time-series datasets, spanning a variety of domains, including medical data, smartwatch sensor data, household power consumption data,  gas sensor array data, etc. Our first two datasets are taken from the WESAD dataset \cite{wesad}. As shown in Table \ref{table1}, WESAD is a publicly available dataset for wearable stress and affect detection. The dataset, named EDA, has data from the electrodermal activity sensor. It is sampled at 4 Hz and the data is provided in $\mu$S. The dataset, named BVP, has data from photoplethysmograph (PPG). It is sampled at 64 Hz. The other three datasets were taken from the UCI Machine Learning Repository \cite{Dua:2019}. These are individual household electric power consumption (sampling once in every minute) \cite{power}, Heterogeneity Human Activity Recognition \cite{10.1145/2809695.2809718} - smartwatch accelerometer and smartwatch gyroscope and Home activity monitoring - MOX gas sensors resistance \cite{HUERTA2016169} (sampling once every second). Table \ref{table1} shows the datasets, their descriptions, and their original sizes. Some of the datasets contained multivariate time series, out of which a single variable was considered for univariate time series compression.

% \subsection{Performance Metrics}

We use compression ratio (CR), compression rate, and decompression rate for comparing the proposed compression methods with the state-of-the-art methods. CR for a file is defined as, CR = {size of file before compression}/{size of file after compression}. CR value is desired to be high. Compression rate ($C_{rate}$) for the file is defined as, ($C_{rate}$)= {size of file before compression}/{time taken}. Similarly, decompression rate ($D_{rate}$) for the file is defined as, ($D_{rate}$)= {size of file before decompression}/{time taken}. The experiments are run on Windows 10 Home laptop with 2.40 GHz Intel processors with Turbo Boost Technology and 16 GB RAM.

% In case for when two or more files are taken, compression ratio is defined as 
% \[
%         CR= \frac{\text{total sum of sizes of all files before compression}}{\text{total sum of size of all files after compression}}
%     \]
\subsection{Performance Evaluation}
In this subsection, we present the simulation results for the proposed version 1 and version 2 methods.   To understand the impact of different entropy schemes on compression ratio, three different entropy encoding schemes (static Huffman, adaptive Huffman and adaptive arithmetic) were applied post indicator function. CR values for version 1 and version 2 methods are shown in Table \ref{table2:v1} and \ref{table2:v2}. Simulations in Table \ref{table2} are shown for different entropy encoders with maximum allowed error of $10^{-3}$. From Table \ref{table2:v1} and \ref{table2:v2}, we see that adaptive arithmetic archives the best compression among other entropy encoders in most cases, except for EDA dataset where static Huffman has bet performance. It can also be seen that while adaptive Huffman may require less resources than static Huffman, static Huffman provides better compression ratio.

\begin{table}[t!]
    \centering
    
    \medskip
    \caption{Compression ratio and compression time (in seconds) taken by statistical method version 1  for lossless case.}
    \label{table3}
    \resizebox{.7\columnwidth}{!}{
    \begin{tabular}{|p{1.5cm}|c|c|c|c|c|c|}
    \hline
    Datasets &  BVP & EDA & ACM & GYS & GAS & Gactive\\
    \hline
    CR & 2.80 & 3.02 & 2.68 & 3.22 & 3.91 & 3.08\\
    \hline
    Time  & 0.385 & 0.027 & 1.703 & 1.741 & 0.993 & 2.008\\
    \hline
    \end{tabular}}
\end{table}
Statistical methods: version 1 and version 2 both could be used for lossless case, i.e., when there is no quantization, the CR performance along with time taken can be seen in Table \ref{table3}. It can be observed that time taken by the algorithm varies between 0 to 2 seconds which is far than time taken to collect the same amount of data. From Table \ref{table2} and \ref{table3}, it can be observed that version 1 and version 2 with maximum error constraint provide significant CR improvement.

\begin{table}[htbp!]
    \centering
    \caption{Performance of version 1 and 2 on GAS and BVP datasets.}
    \label{table5}
    \resizebox{.7\columnwidth}{!}{
    \begin{subtable}[h]{0.5\columnwidth}
    \centering
    \begin{tabular}{|c|c c c|}
    \hline
    \multirow{2}{*}{Version} & \multicolumn{3}{|c|}{Block Size}\\  & 16 & 32 & 64\\
    \hline
    1    &  13.84 & 12.91 & 11.23\\
    \hline
    2    &  14.37 & 15.69 & 22.07\\
    \hline
    \end{tabular}
    \caption{GAS}
    \label{table5:v1}
    \end{subtable}
    \begin{subtable}[h]{0.5\columnwidth}
    \centering
    \begin{tabular}{|c|c c c|}
    \hline
    \multirow{2}{*}{Version} & \multicolumn{3}{|c|}{Block Size}\\  & 16 & 32 & 64\\
    \hline
    1    &  2.79 & 2.83 & 2.85\\
    \hline
    2    &  2.43 & 2.54 & 2.59\\
    \hline
    \end{tabular}
    \caption{BVP}
    \label{table5:v2}
    \end{subtable}}
    
\end{table}
 
\medskip
\begin{table}[htbp!]

\caption{Compression ratio vs $\tau$ for block size 16 with maximum error of $10^{-3}$.}
\label{table6}
% \begin{subtable}[h]{0.49\textwidth}
% \centering
% \begin{tabular}{| p{1cm} | p{1.5cm} | p{1.5cm} | p{1.5cm} |} 
%  \hline
% \multirow{2}{*}{Datasets} & \multicolumn{3}{|c|}{Tau($\tau$) values}\\ \cline{2-4} & 5 & 7 & 9\\
% \hline
% BVP & \textbf{2.80} & 2.80 &	2.80\\
% \hline
% EDA	& \textbf{12.59} & 12.56 & 12.20\\
% \hline
% ACM & \textbf{5.44} & 5.42 & 5.42\\
% \hline
% GYS & 7.82 & \textbf{7.85} & 7.83\\
% \hline
% GAS & 13.74 & \textbf{13.84} & 13.63 \\
% \hline
% Gactive & 4.17& \textbf{4.18} & 4.14\\
% \hline

% \end{tabular}
% \caption{Version 1}
% \label{table6:v1}
% \end{subtable}
% \begin{subtable}[h]{0.49\textwidth}
\centering
\resizebox{.7\columnwidth}{!}{
\begin{tabular}{| p{1cm} | p{1.5cm} | p{1.5cm} | p{1.5cm} |} 
 \hline
\multirow{2}{*}{Datasets} & \multicolumn{3}{|c|}{Tau($\tau$) values}\\ \cline{2-4} & 5 & 7 & 9\\
\hline
BVP & 2.44 & 2.44 & \textbf{2.44}\\
\hline
EDA & 12.52 & 12.66 & \textbf{12.75}\\
\hline
ACM & \textbf{4.69} & 4.65 & 4.64\\
\hline
GYS & 6.95 & 7.04 & \textbf{7.13}\\
\hline
GAS & 13.80 & 14.37 & \textbf{15.10}\\
\hline
Gactive	& 3.99 & 4.10 & \textbf{4.18}\\

\hline
\end{tabular}}
%\caption{Version 2}
%\label{table6:v2}
%\end{subtable}
\end{table}

\medskip
\begin{table}[htbp!]

\medskip
\caption{Compression ratio vs $\tau$ for block size 32 with maximum error of $10^{-3}$.}
\label{table7}
% \begin{subtable}[h]{0.49\textwidth}
% \centering
% \begin{tabular}{| p{1cm} | p{1cm} | p{1cm} | p{1cm} | p{1cm} |} 
%  \hline
% \multirow{2}{*}{Datasets} & \multicolumn{4}{|c|}{Tau($\tau$) values}\\ \cline{2-5} & 5 & 9 & 13 & 17\\
% \hline
% BVP & 2.78 & 2.78 & 2.78 & \textbf{2.78}\\
% \hline
% EDA	& 12.06 & 12.20 & 12.12 & \textbf{11.94}\\
% \hline
% ACM & \textbf{5.44} & 5.43 & 5.42 & 5.42\\
% \hline
% GYS & 7.60 & 7.62 & \textbf{7.63} & 7.61\\
% \hline
% GAS & 12.23 & 12.49 & 12.62 & \textbf{12.67} \\
% \hline
% Gactive & 3.85 & 3.98 &	\textbf{4.05} & 4.03\\
% \hline

% \end{tabular}
% \caption{Version 1}
% \label{table7:v1}
% \end{subtable}
% \begin{subtable}[h]{0.49\textwidth}
\centering
\resizebox{.7\columnwidth}{!}{
\begin{tabular}{| p{1cm} | p{1cm} | p{1cm} | p{1cm} | p{1cm} |} 
 \hline
\multirow{2}{*}{Datasets} & \multicolumn{4}{|c|}{Tau($\tau$) values}\\ \cline{2-5} & 5 & 9 & 13 & 17\\
\hline
BVP & 2.51 & 2.51 &	2.51 & \textbf{2.51}\\
\hline
EDA & 11.90 & 12.36 & 12.86 & \textbf{13.29}\\
\hline
ACM & \textbf{4.95} & 4.87 & 4.86 & 4.86\\
\hline
GYS & 6.89 & 7.05 & 7.23 & 	\textbf{7.40}\\
\hline
GAS & 12.10 & 12.83 & 13.96 & \textbf{15.55}\\
\hline
Gactive	& 3.77 & 4.06 & 4.36 & \textbf{4.47}\\

\hline
\end{tabular}}
%\caption{Version 2}
%\label{table7:v2}
%\end{subtable}
\end{table}

\begin{table}[htbp!]

\medskip
\caption{Compression ratio vs $\tau$ for block size 64 with maximum error of $10^{-3}$.}
\label{table8}
% \begin{subtable}[h]{0.49\textwidth}
% \centering
% \begin{tabular}{| p{1cm} | p{1cm} | p{1cm} | p{1cm} | p{1cm} |} 
%  \hline
% \multirow{2}{*}{Datasets} & \multicolumn{4}{|c|}{Tau($\tau$) values}\\ \cline{2-5} & 10 & 20 & 30 & 40\\
% \hline
% BVP & 2.80 & 2.80 & 2.80 & 2.80\\
% \hline
% EDA	& 11.14 & 11.49 & \textbf{11.59} & 11.09\\
% \hline
% ACM & \textbf{5.48} & 5.47 & 5.47 & 5.47\\
% \hline
% GYS & 7.40 & 7.54 & \textbf{7.58} & 7.50\\
% \hline
% GAS & 10.11 & 10.60 & \textbf{11.16} & 11.16\\
% \hline
% Gactive & 3.80 & 4.04 & \textbf{4.06} & 4.05\\
% \hline

% \end{tabular}
% \caption{Version 1}
% \label{table8:v1}
% \end{subtable}
%\begin{subtable}[h]{0.49\textwidth}
\centering
\resizebox{.7\columnwidth}{!}{
\begin{tabular}{| p{1cm} | p{1cm} | p{1cm} | p{1cm} | p{1cm} |} 
 \hline
\multirow{2}{*}{Datasets} & \multicolumn{4}{|c|}{Tau($\tau$) values}\\ \cline{2-5} & 10 & 20 & 30 & 40\\
\hline
BVP & 2.56 & 2.56 & 2.56 & 2.56\\
\hline
EDA & 11.06 & 12.18 & 13.05 & \textbf{14.04}\\
\hline
ACM & \textbf{5.02} & 4.99 & 4.99 & 4.99\\
\hline
GYS & 6.93 & 7.26 & 7.56 & \textbf{7.74}\\
\hline
GAS & 10.13 & 11.63 & 14.57 & \textbf{19.17}\\
\hline
Gactive	& 3.82 & 4.46 & 4.66 & \textbf{4.69}\\

\hline
\end{tabular}}
%\caption{Version 2}
\label{table8:v2}
%\end{subtable}
\end{table}

 Empirically, we observed that the version 1 performs better when the data is fluctuating and there is no consecutive increasing or decreasing pattern like BVP dataset. Whereas, version 2 performs better when there is a trend (increasing or decreasing) over a period of time like ACM dataset.

The effect of changing the block size of for version 1  and version 2 with maximum error of $10^{-3}$ on GAS and BVP datasets can be seen from Table. \ref{table5}. It shows both versions of statistical methods behave differently on increasing the block size. Compression ratio of version 1 decreases while the compression ratio of version 2 increases on increasing the block size. This property for version 1 tends to differ with different dataset. For datasets with fluctuating values compression ratio remains constant or increases very slowly like in BVP. 

$\tau$ is another variable whose alteration will have an effect on the compression ratio. Tables \ref{table6}, \ref{table7}, and \ref{table8} show how the compression ratio changes with different $\tau$ values and different block sizes for version 2. It was observed in version 1 that either the values are constant or increase rather slowly till a particular value. Compression ratio for version 2 either remains constant or are sharply increasing with increase in value of $\tau$. This difference in trend can be explained by the fact that version 2 applies IFZ function in case of difference coding, whereas version 1 does not. The simulation for version 1 is available at \url{https://github.com/vidhi0206/data-compression}.

% \begin{table}[t!]
%     \centering
%     \caption{Compression Ratio in cases where datasets are taken individually before entropy encoding and when datasets are clubbed together before entropy encoding. }
%     \label{aggregator}
%     \begin{tabular}{|c|c|c|}
%     \hline
%         Datasets &  taken individually & combined together  \\
%     \hline
%         BVP + EDA & \textbf{3.00} & 2.82\\
%         \hline
%         Gactive + GAS & \textbf{7.00} & 6.87\\
%         \hline
%         ACM + GYS & 6.00 & \textbf{6.30}\\
%         \hline
%     \end{tabular}
% \end{table}
% As mentioned in Section \ref{comp-method} we can also use aggregator to combine blocks of different datasets with similar values together before applying entropy encoding. Table \ref{aggregator} shows the compression ratio attained in case of without and with combining datasets before entropy encoding. ACM and GYS provide better combined compression ratio when aggregator is used, wheresas other datasets have better combined compression ratio when they are taken individually before applying the entropy encoding.

% Looking at the above results, we can see that to obtain maximum compression allowed by the algorithm for a particular dataset, pattern of data needs to be observed carefully. Optimal values of $\tau$ and block size need to be experimentally obtained for high compression ratio.

 \begin{table}[htbp!]
     \centering
     
     \medskip
     \caption{ CR comparison for CA, SZ, LFZip (NLMS) and the proposed statistical method version 2.}
     \label{table4}
     
     \begin{subtable}[h]{0.48\textwidth}
    \centering
    \resizebox{.7\columnwidth}{!}{
     \begin{tabular}{| p{1cm} | p{1.5cm} | p{1cm}  p{1cm}  p{1cm} |}
     \hline
    \multirow{2}{*}{Datasets} & \multirow{2}{*}{Compressor} & \multicolumn{3}{|c|}{Maximum Error}\\
    & & $10^{-3}$ & $10^{-2}$ & $10^{-1}$\\
    \hline
    \multirow{4}{*}{BVP} & CA & 2.48 & 2.49 & 2.74\\ 
        & SZ & 2.43 & 2.80 & 4.39\\
        & LFZip & \textbf{3.18} & \textbf{5.28} & \textbf{9.13}\\
        & Version 2 & 2.80 & 3.30 & 7.03\\
    \hline
     \multirow{4}{*}{ACM} & CA & 2.84 & 3.10 & 5.19\\ 
        & SZ & 3.25 & 5.05 & 11.00\\
        & LFZip & 3.55 & 5.86 & 12.71\\
        & Version 2 & \textbf{5.06} & \textbf{9.04} & \textbf{38.79}\\
    \hline
     \end{tabular}}
    \label{table4:p1}
     \end{subtable}
    \begin{subtable}[h]{0.48\textwidth}
    \centering
    \resizebox{.7\columnwidth}{!}{
     \begin{tabular}{| p{1cm} | p{1.5cm} | p{1cm}  p{1cm}  p{1cm} |}
     \hline
    % \multirow{2}{*}{Datasets} & \multirow{2}{*}{Compressor} & \multicolumn{3}{|c|}{Maximum Error}\\
    % & & $10^{-3}$ & $10^{-2}$ & $10^{-1}$\\
    % \hline
    \multirow{4}{*}{GYS} & CA & 2.88 & 4.27 & 10.75\\ 
        & SZ & 4.26 & 8.08 & 24.79\\
        & LFZip & 6.05 & 12.26 & 28.77\\
        & Version 2 & \textbf{7.95} & \textbf{19.59} & \textbf{100.94}\\
    \hline
     \multirow{4}{*}{Gactive} & CA & 5.05 & 6.23 & 12.47\\ 
        & SZ & 5.09 & 9.65 & 23.99\\
        & LFZip & 4.17 & 7.37 & 17.98\\
        & Version 2 & \textbf{5.06} & \textbf{9.04} & \textbf{38.79}\\
    \hline
     \end{tabular}}
    \label{table4:p2}
     \end{subtable}
 \end{table}
\begin{figure}[htbp]
  \includegraphics[width=\linewidth]{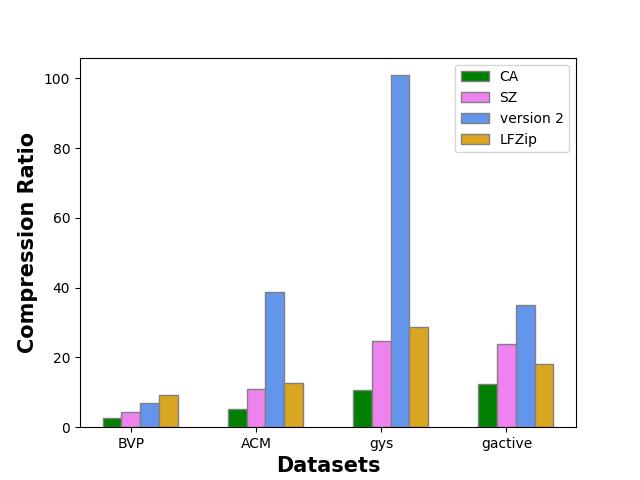}
  \caption{Compression performance on several data sets with the proposed statistical method (version 2). }
  \label{comparision}
\end{figure}
\subsection{Performance Comparison} From previous subsection, it is clear that version 2 performs better than version 1 for most cases. Therefore, in this subsection, we compare version 2 method with the state-of-art methods. 
Performance of the proposed version 2 is compared  with LFZip \cite{9105816}, CA \cite{7076402} and SZ \cite{7516069} compression methods. LFZip uses NLMS predictor followed by BSC, an efficient BWT-based compressor. CA uses static curve fitting model and transmits data that could not fit the model. It is used in industry for low computational requirements. SZ uses a dynamic curve fitting model that adapts to the data  that could not fit the model.  It is the current state-of-the-art compressor for maximum error distortion. Table \ref{table4} shows the results for CA, SZ, LFZip (using NLMS predictor with default window size k = 32) and statistical method version 2 (using block size = 64.) for four univariate time series datasets and three values of maximum error ($10^{-1}, 10^{-2}, 10^{-3}$). We can see that, statistical method (version2) performs better in every case except BVP. It can be seen that having a higher level of distortion helps us get a better compression ratio. The compression archived by the state-of-the-art methods  and the proposed version 2 are compared in Fig. \ref{comparision}. As seen from Fig. \ref{comparision} statistical method (version 2) performs better than that of other methods in most cases, except of BVP dataset where LFZip compressor performs better. Statistical method in Fig. \ref{comparision} is simulated with maximum error as $10^{-1}$ and block size 64 and $\tau = 50$.
  
  \begin{table}[htbp!]
      \centering
      
      \caption{Encoding and decoding rate (MB/seconds) for version 2 with block size 16.}
      \label{table9}
      \resizebox{.7\columnwidth}{!}{
      \begin{tabular}{|p{1.1cm}|p{.6cm}|p{.6cm}|p{.6cm}|p{.6cm}|p{.6cm}|p{.8cm}|}
      \hline
       Datasets & BVP & EDA & ACM & GYS & GAS & Gactive\\
       \hline
       Encoding Rate & 0.692 & 1.181 & 1.299 & 1.600 & 2.829 & 1.011\\
        \hline
        Decoding Rate & 0.630 & 1.436 & 1.141 & 1.222 & 2.870 & 0.982\\
        \hline
      \end{tabular}}
  \end{table}
  
 Table \ref{table9} presents the compression/decompression rate achieved by the algorithm. Both compression and decompression rates varies between 0.6 and 2.8 MB/sec. The rate is slower than LFZip (NLMS) which has the rate about 2M timesteps/s \cite{9105816} but should be practical for most application. 

\section{Conclusion}
\label{conclusion}
We have proposed time-series data compression methods using statistical and difference encoding. The proposed methods are compared with the state-of-the-art compression methods, i.e., LFZip, CA, and SZ. From simulation results, it is observed that one of the proposed compression methods provides 2x improvement in compression ratio for ACM and GYS datasets with akin compression  and decompression rate as compared to the state-of-the-art compression methods. 
In the future, we plan to extend the proposed compression methods for multi-dimensional multi-sensor time series data. 
      
\subsection*{Authorship contribution}
Vidhi Agrawal simulated, modified, fine-tuned compression methods, and wrote the paper. Gajraj Kuldeep proposed  compression methods and participated in supervision.  Dhananjoy Dey supervised the overall work.

\printbibliography[
 heading=bibintoc,
 title={References}]

\end{document}